\documentclass[12pt]{article}

\usepackage{scicite}
\usepackage{graphicx}
\usepackage{times}
\usepackage{multirow}
\usepackage{color}

\newenvironment{sciabstract}{%
\begin{quote} \bf}
{\end{quote}}

\title{Measuring the quantum state of photoelectrons}

\author
{Hugo Laurell,$^{1,\dag}$ Sizuo Luo,$^{1,\dag}$ Robin Weissenbilder$^{1}$, Mattias Ammitzböll$^{1}$, \\
Shahnawaz Ahmed$^{2}$, Hugo S\"{o}derberg$^{1}$, C. Leon. M. Petersson$^{3}$,V\'enus Poulain$^{1}$, \\
Chen Guo$^{1}$, Christoph Dittel$^{4,5,6}$, Daniel Finkelstein-Shapiro$^{7}$, \\ Richard J. Squibb$^{8}$, Raimund Feifel$^{8}$, Mathieu Gisselbrecht$^{1}$, \\
Cord L. Arnold$^{1}$, Andreas Buchleitner$^{4,5}$, Eva Lindroth$^{3}$,  \\ 
Anton Frisk Kockum$^{2}$, Anne L'Huillier$^{1}$, David Busto$^{1,*}$\\
\\
\normalsize{$^{1}$Department of Physics, Lund University, Box 118, 221 00 Lund, Sweden}\\
\normalsize{$^{2}$Department of Microtechnology and Nanoscience,}\\
\normalsize{Chalmers University of Technology, 412 96 Gothenburg, Sweden}\\
\normalsize{$^{3}$Department of Physics, Stockholm University,}\\
\normalsize{ AlbaNova University Center, 106 91 Stockholm, Sweden}\\
\normalsize{$^{4}$Physikalisches Institut, Albert-Ludwigs-Universit\"at Freiburg,}\\
\normalsize{Hermann-Herder-Straße 3, 79104 Freiburg, Germany}\\
\normalsize{$^{5}$EUCOR Centre for Quantum Science and Quantum Computing,}\\
\normalsize{Albert-Ludwigs-Universit\"at Freiburg, Hermann-Herder-Straße 3,}\\
\normalsize{79104 Freiburg, Germany}\\
\normalsize{$^{6}$Freiburg Institute for Advanced Studies, Albert-Ludwigs-Universit\"{a}t Freiburg,}\\
\normalsize{Albertstraße 19, 79104 Freiburg, Germany}\\
\normalsize{$^{7}$Instituto de Qu\'{ı}mica, Universidad Nacional Aut\'{o}noma de M\'{e}xico, Circuito Exterior,}\\
\normalsize{Ciudad Universitaria, Alcald\'{ı}a Coyoac\'{a}n C.P. 04510, Ciudad de M\'{e}xico, Mexico}\\
\normalsize{$^{8}$Department of Physics, University of Gothenburg, Origov\"{a}gen 6B,}\\
\normalsize{412 96 Gothenburg, Sweden}\\
\\
\normalsize{$^\ast$ Corresponding author; E-mail:   david.busto@fysik.lth.se}\\
\normalsize{$^\dag$ These authors contributed equally to this work}\\
}

\date{}

\begin{document} 


\baselineskip24pt


\maketitle

\begin{sciabstract}
A photoelectron, emitted due to the absorption of light quanta as described by the photoelectric effect, is often characterized experimentally by a classical quantity, its momentum. However, since the photoelectron is a quantum object, its rigorous characterization requires the reconstruction of the complete quantum state, the photoelectron's density matrix. Here, we use quantum state tomography to fully characterize photoelectrons emitted from helium and argon atoms upon absorption of ultrashort, extreme ultraviolet light pulses. While in helium we measure a pure photoelectronic state, in argon, spin-orbit interaction induces entanglement between the ion and the photoelectron, leading to a reduced purity of the photoelectron state. Our work shows how state tomography gives new insights into the fundamental quantum aspects of light-induced electronic processes in matter, bridging the fields of photoelectron spectroscopy and quantum information, and offering new spectroscopic possibilities for quantum technology.

\end{sciabstract}

\paragraph*{Introduction}

The explanation of the photoelectric effect by Einstein in his \textit{annus mirabilis} in 1905~\cite{Einstein1905} was an important catalyst for the development of quantum mechanics. 
The photoelectric effect occurs when atoms, molecules, or solids absorb high-energy photons, resulting in the emission of electrons. Today, this fundamental quantum effect underpins various modern measurement techniques involving electrons, such as microscopy\cite{Swiech1997,Dabrowski2020}, holography\cite{FigueiraMorissonFaria2020}, diffraction\cite{Woodruff2002}, and spectroscopy\cite{Becker1995,Stolow2003}. These techniques rely on measuring the photoelectron's kinetic energy or momentum to gain information about the structural and chemical properties of matter.
The advent of attosecond science has led to a paradigm shift in experiments by giving access to both the modulus and the spectral phase variation of the photoelectron momentum distribution~\cite{Hentschel2001,Paul2001}. Spectral phase measurements reveal the existence of attosecond delays in photoionization~\cite{Schultze2010,Klunder2011}; this sensitivity to small details of the ionic potential has been used to gain unprecedented insights into electron correlations \cite{Alexandridi2021, Gruson2016, Zhong2020}, electron-nuclear couplings in molecules~\cite{Cattaneo2018,Cattaneo2022}, the influence of the molecular structure on the photoionization dynamics~\cite{Biswas2020}, as well as the initial state localization and electron transport in solids \cite{Locher2015,Siek2017,Ossiander2018}.

Despite the undeniable success of existing attosecond photoelectron interferometry techniques, 
when the photoelectron spectra result from the incoherent addition of different transition probabilities, the phase extracted from the interferograms cannot be easily and unambiguously related to those of individual ionization channels~\cite{Isinger2017,Alexandridi2021,Peschel2022,Busto2022}. In other words, these techniques work well only if the photoelectron is properly described by a pure state, while the characterization of mixed photoelectron states remains a problem. The presence of mixed photoelectron states, however, is the general rule,  since entanglement between different degrees of freedom is common in photoionization processes~\cite{Pabst2011,Nishi2019,Carlstrom2018,Vrakking2021,Koll2022,Busto2022}. When all involved degrees of freedom cannot be simultaneously measured (e.g., the ion may not be addressable), this entanglement induces a mixed reduced state of the photoelectron, which is referred to as decoherence.  In this case, quantum state tomography (QST) is required to fully characterize the photoelectron quantum state.
 Two methods have been experimentally investigated for QST of electrons in the continuum: Mixed-FROG (frequency-resolved optical gating for mixed states)~\cite{Bourassin2015} and SQUIRRELS (spectral quantum interference for the regularized reconstruction of free-electron states)~\cite{Priebe2017}. The former method (Mixed-FROG) was used to characterise photoelectrons emitted from neon by absorption of attosecond pulse trains~\cite{Bourassin2020}. In that work, the low degree of coherence of the reconstructed quantum state was attributed to experimental imperfections and referred to as experimental decoherence. The latter method (SQUIRRELS) aims at characterizing attosecond electron pulse trains in electron transmission microscopy, neglecting the continuous nature of the electron kinetic energy. This highlights that establishing photoelectron QST as a new tool for the investigation of entanglement and decoherence in atomic and molecular processes requires a very low degree of experimental decoherence and a high spectral resolution in order to measure subtle differences between photoelectrons originating from different targets. So far, these conditions have not been met.
 
In this work, we tomographically reconstruct the quantum state of a photoelectron, under conditions of very low experimentally induced decoherence, via the KRAKEN (Swedish acronym standing for quantum state tomography of attosecond electron wavepacket) method proposed in Ref.~\cite{Laurell2022}. We photoionize helium and argon atoms using short extreme ultraviolet (XUV) pulses, populating a broad superposition of continuum states. We  reconstruct the density matrix of the photoelectron emitted after absorption of a single XUV photon using a delayed, spectrally tunable, bichromatic infrared (IR) pulse as a probe~(Fig. 1). 
By varying the frequency separation of the two spectral components of the probe pulse, we address different off-diagonal elements, allowing us to reconstruct the full density matrix of the photoelectron. In helium, we measure an almost perfectly pure photoelectronic state, while in argon, we measure a mixed state. The obtained purity is consistent with the reduction expected due to entanglement between the ion and the photoelectron, induced by spin-orbit interaction. 

\paragraph*{Principle of the experiment}
Figure~1 schematically presents the principle of the experiment. A femtosecond XUV pulse with 30~eV photon energy, obtained via high-order harmonic generation in a gas cell, ionizes the target, populating a broad superposition of continuum states of energy $\epsilon$. The total quantum system describing the combined system of photoelectron and ion has a large number of degrees of freedom, which, in most cases, cannot be accessed simultaneously. In our experiment, we address the photoelectron and determine its reduced quantum state by measuring its kinetic energy using a magnetic bottle electron spectrometer [Fig.~1(B)]. The population of the photoelectron quantum state, given by the diagonal elements of the density matrix, $\rho(\epsilon,\epsilon)=\langle \epsilon|\rho|\epsilon\rangle$, can, in principle, be obtained by measuring the photoelectron spectrum. In order to measure the coherences $\rho(\epsilon_1,\epsilon_2)=\langle \epsilon_1|\rho|\epsilon_2\rangle$ between two continuum states $|\epsilon_1\rangle$ and $|\epsilon_2\rangle$, we probe the photoelectron with a bichromatic IR pulse with synchronized and phase-locked spectral components (angular frequencies $\omega_1$ and $\omega_2$) \cite{Laurell2022}, which induces an additional photon absorption to a final state $|\epsilon_f\rangle$ with energy
\begin{equation}
 \epsilon_f=\epsilon_1+\hbar\omega_1=\epsilon_2+\hbar\omega_2.   
\end{equation}
 When the delay $\tau$ between the pump (XUV) and probe (IR) pulses is varied, the photoelectron yield in the final state oscillates at a frequency $\delta\omega=\omega_1-\omega_2$ (i.e., a beating between the two spectral IR components). Extracting the amplitude $A_{\delta\omega}(\epsilon_f)$ and phase $\phi_{\delta\omega} (\epsilon_f)$ of these oscillations allows us to determine one off-diagonal element of the density matrix according to $A_{\delta\omega}(\epsilon_f)\propto|\rho(\epsilon_1,\epsilon_2)|$ and $\phi_{\delta\omega}(\epsilon_f)=\arg[\rho(\epsilon_1,\epsilon_2)]$. By measuring these quantities as a function of the final photoelectron kinetic energy $\epsilon_f$, a subdiagonal of the density matrix, horizontally shifted from the main diagonal by $\hbar\delta\omega$, is obtained~\cite{Laurell2022}. The density matrix is then determined by repeating these measurements for different values of the frequency separation between the two IR spectral components. In our measurements, the frequency $\omega_1$ is fixed and only $\omega_2$ is varied [see Fig. 1(C)]. Details of the setup used for the experiments can be found in the supplementary material (SM) and in Ref.~\cite{Luo2023}.

\paragraph*{Quantum state tomography of a pure photoelectronic state}

Figure ~\ref{Fig2}(A) presents the photoelectron spectra obtained by ionizing helium as a function of delay $\tau$ for seven values of $\delta\omega$. The measurement for $\delta\omega=0$ gives access to the populations (the diagonal elements in the density matrix $\rho$). Figure~\ref{Fig2}(B) shows, for each $\delta\omega$, the oscillation amplitude $A_{\delta\omega}(\epsilon_f)$ as a function of the final kinetic energy. The amplitude of the oscillations shows a clearly decreasing trend with increasing $\delta\omega$. In our measurements, the phase of the oscillations only carries information about the properties of the classical ionizing XUV field. The phase of the density matrix is presented and discussed in the SM.

The photoelectron density matrix can be reconstructed by inserting each measured amplitude $A_{\delta\omega}$ at the corresponding subdiagonal of an initially empty density matrix. Figure 2(C) presents the density matrix after measuring a finite number of subdiagonals, scaled by a global factor to ensure a unit trace and mirrored with respect to the diagonal (see SM). The amplitude of the density matrix describes an approximately circular pattern. The dark areas in the density matrix correspond to elements that are not covered by the finite number of measured subdiagonals.

Measuring the full quantum state of photoelectrons is a nontrivial problem  due to the continuous nature of the photoelectron energy distribution. This problem is similar to that encountered in quantum-optics experiments aiming, for example, at measuring the density matrix of squeezed quantum states of light~\cite{Smithey1993}. Different methods such as maximum-likelihood or maximum-entropy reconstructions have been developed to estimate the quantum state based on a finite number of measurements~\cite{Lvovsky2009}, and machine-learning techniques have recently been applied to QST~\cite{Gebhart2023}. Here we employ Bayesian estimation using a Hamiltonian Monte Carlo method to extract the density matrix from our measurements (more details are presented in the SM). The results are presented in Fig.~\ref{Fig3}(A). The retrieved density matrix is approximately circular. However, a slight elongation along the diagonal can be observed, with the coherence between continuum states with a large energy separation being reduced. It can be shown that this effect is due to the finite resolution of the electron spectrometer~\cite{Mucke2012}, which leads to a decrease in the measured photoelectron purity. We compensate for this limitation by measuring the spectrometer response and feeding the result to the Bayesian estimation algorithm (more details in the SM).  The new reconstructed quantum state is presented in Fig.~3(B). The amplitude of the density matrix has now an almost perfectly circular shape, indicating that the photoelectron is described by an essentially pure coherent state (see SM for the phase of the density matrix).

We compare our result to theoretical calculations based on the relativistic random phase approximation with exchange (RRPAE) \cite{Vinbladh2022} (details in the SM). Figure~3{(C)} presents the theoretical density matrix obtained from two-photon (XUV-IR) calculations, following a similar procedure as in the experiment, without Bayesian optimization. The experimental (B) and theoretical (C) results are in excellent agreement. As a final step, Fig.~3(D) presents the result of a calculation of the density matrix of a photoelectron emitted by absorption of the XUV radiation only (hence the kinetic energy scale, which differs by the energy of one IR photon, 1.55 eV). The photoelectron quantum state is almost identical to the density matrices obtained from the experimental and theoretical KRAKEN reconstructions. We quantify the agreement between the amplitudes of the experimental and the XUV-only theoretical density matrices, $\rho_\mathrm{exp}$ and $\rho_\mathrm{theo}$,  by evaluating their mutual fidelity \cite{Uhlmann2009}, $F(\rho_\mathrm{exp},\rho_\mathrm{theo})=\mathrm{tr}\left[(r_\mathrm{theo}
^{1/2}r_\mathrm{exp}r_\mathrm{theo}^{1/2})^{1/2}\right]=0.98$, where $r=|\rho|$. This result demonstrates that our protocol accurately measures the quantum state of the photoelectron emitted by the XUV pulse, in agreement with previous theoretical work~\cite{Laurell2022}.

\paragraph*{Quantum state tomography of a mixed photoelectronic state}%

We now present measurements of the quantum state of photoelectrons emitted from argon atoms excited at the same photon energy. Similar to our measurements in helium, a series of seven spectrograms, one for each value of $\delta\omega$, is acquired while keeping all other experimental parameters unchanged. Contrary to helium, in the case of argon, spin-orbit interaction splits the ionic ground state, leading to two possible final states, 3p$^5\ ^2$P$_{3/2}$ and 3p$^5\ ^2$P$_{1/2}$, separated by $\Delta\epsilon_\textrm{so}\approx 177$~meV. As a result, and as schematically shown in Fig.~4(A), the photoelectron spectrum is composed of two photoelectron peaks shifted in energy by $\Delta\epsilon_\textrm{so}$. 
Figure~4(B) presents the photoelectron density matrix reconstructed from the experimental data (raw data are shown in the SM). Similarly to our helium measurements, the finite spectrometer resolution is accounted for in the reconstruction. Compared to the quantum state measured with helium, in the case of argon, the density matrix exhibits a strong elongation along the diagonal. This indicates a loss of coherence between continuum states with large energy separation. 

The degree of coherence of the photoelectron quantum state can be quantified by calculating its purity $\gamma=\textrm{tr}(\rho^2)$,  with $\gamma=1$ corresponding to a fully coherent state. Table~\ref{tab:table1} presents the purity of the reconstructed photoelectron density matrices in helium and argon, and compares it to the purity predicted by RRPAE calculations of single-photon ionization. In helium, where spin-orbit interaction is absent, the purity is close to unity, $\gamma=0.94\pm0.06$, demonstrating a nearly perfectly pure coherent state as well as practically no experimental decoherence besides that induced by the spectrometer response function. In contrast, the purity of the quantum state of the photoelectron originating from argon is reduced, $\gamma=0.65\pm0.02$, and nearly agrees with the theoretical prediction of $\gamma=0.61$. 
The small discrepancy in argon might originate from small fluctuations of the XUV spectral width during the measurements.

\begin{table}[h]
\centering
\begin{tabular}{cccc}
\hline
\multicolumn{2}{|c|}{}                                                   & \multicolumn{1}{c|}{helium} &  \multicolumn{1}{c|}{argon} \\ \hline
\multicolumn{1}{|c|}{\multirow{2}{*}{~Purity~}} & \multicolumn{1}{c|}{~Experiment~} & \multicolumn{1}{c|}{~$0.94\pm0.06$~}  &  \multicolumn{1}{c|}{~$0.65\pm0.02$~}  \\ \cline{2-4} 
\multicolumn{1}{|c|}{} & \multicolumn{1}{c|}{Theory}  & \multicolumn{1}{c|}{1.00}  &  \multicolumn{1}{c|}{0.61}  \\ \hline
\end{tabular}
    \caption{Experimentally and theoretically extracted purity of the reduced photoelectron quantum state for helium and argon. The experimental uncertainties correspond to the 95\% confidence interval (see SM). The theoretical values correspond to the purity of the photoelectron quantum state after absorption of XUV-only radiation.}
    \label{tab:table1}
\end{table}

 The bipartite ion+photoelectron system can be considered to be fully coherent because, due to the nature of the generation process, the ionizing XUV field is coherent. Additionally, the time scale involved in the preparation and characterization of the photoelectron quantum state is in the 100 fs range, which is significantly shorter than the typical time scale for interactions with the environment. 
The state of the ion+photoelectron system in argon is therefore given by \cite{Laurell2022}
\begin{equation}
        |\Psi\rangle=\ \frac{1}{\sqrt{3}}|\phi_{1/2}\rangle\otimes|\psi_{1/2}\rangle+\sqrt{\frac{2}{3}}|\phi_{3/2}\rangle\otimes|\psi_{3/2}\rangle,
\label{Eq:TotalSystem}
\end{equation}
where the subscript $j=\{1/2,3/2\}$ refers to the ionic angular momentum, $|\phi_{j}\rangle$ is the ionic wavefunction, with $\langle \phi_j | \phi_k \rangle = \delta_{j,k}$, and $|\psi_j\rangle$ is the photoelectronic wavefunction. The photoelectronic states, $|\psi_{1/2} \rangle$ and $|\psi_{3/2} \rangle$, corresponding to mutually orthogonal ionic states are distinct since they are centered at different energies (Fig.~\ref{Fig4}A). Accordingly, the ion and the photoelectron are to some degree entangled~\cite{Laurell2022}. Consequently, by interrogating the photoelectron alone, the ion+photoelectron entanglement leads to mixedness of the reduced density matrix of the photoelectron, i.e., a purity smaller than one. 

Pionieering experiments already identified fingerprints of entanglement in photoionization, whether it is between the photoelectron and the ion \cite{Koll2022} (also see discussion in \cite{Tichy2011}), the angular and radial degrees of freedom of the photoelectron \cite{Busto2022}, two photoelectrons in single-photon double ionization \cite{Akoury2007} or fragments in dissociative molecular photoionization \cite{Shobeiry2021}. A crucial advantage of our present method is that, giving access to the photoelectron’s reduced density matrix $\rho$, we are able to 
 quantify the degree of entanglement between the involved degrees of freedom. Since the ionic core of argon lives in an effectively two-dimensional space, the core's and the photoelectron’s joint state has maximally two non-vanishing Schmidt coefficients, and their entanglement can thus be  measured by the concurrence $C=\sqrt{2[1-\textrm{tr}(\rho^2)]}$ \cite{Laurell2022,Tichy2011}. Note that $C$ is a basis independent quantity, and ranges from $C=0$, for a separable state of ion and photoelectron, to $C=1$, for a maximally entangled state of both degrees of freedom. In argon, we obtain a concurrence of $C_\textrm{exp}= 0.84\pm 0.02$, slightly below the theoretical prediction of $C_\textrm{theo}=0.88$.

\paragraph*{Conclusion}
In summary, we performed photoelectron quantum state tomography in helium and argon atoms with a method that enabled us to quantify the influence of ion+electron entanglement on the purity of the photoelectronic quantum state. The demonstrated method can be applied to more complex systems such as molecules, where the measurement of the photoelectron density matrix can provide information on the electronic and nuclear dynamics in the ion \cite{Cattaneo2022,Koll2022,Blavier2022,Blavier2022_2}. 
Applied to the condensed phase, our technique could provide information on the decoherence mechanisms due to interactions with the environment \cite{Pazourek2015,Jordan2020,Silva2022}, possibly with spatial resolution using electron microscopy\cite{Priebe2017}. Finally, our QST protocol could provide a new way to investigate quantum optical effects in high-order harmonic generation (for example squeezing), complementary to the approaches of Refs.~\cite{Gorlach2020,Gorlach2023,Stammer2023,Stammer2022}.

\bibliography{scibib}

\bibliographystyle{Science}


\paragraph*{Funding}
 This work was supported by the Swedish Research Council, the European Research Council (Advanced grant QPAP), and the Knut and Alice Wallenberg Foundation. D.F.S acknowledges funding from PAPIIT No. IA202821. C.D. acknowledges the Georg H. Endress Foundation for support and the Freiburg Institute for Advanced Studies for a FRIAS Junior Fellowship. S. A., A. F. K., M. A. and A.L. acknowledge support from the Knut and Alice Wallenberg Foundation through the Wallenberg Centre for Quantum Technology (WACQT).
 
\paragraph*{Author contributions}
H.L., S.L., R.W., M.A., V.P., C.G. and D.B. designed and constructed the beamline. R.J.S and R.F provided the MBES and helped with its optimization. H.L., S.L., R.W. and M.A. performed the experiments. H.L. analyzed the data with the help of S.L. and D.B.. S.A., A.F.K. and H.S. developed the code for the Bayesian estimation. C.L.M.P and E.L. performed the RRPA calculations. All authors discussed and approved the results and the manuscript. A.L. and D.B. supervised the work and wrote the manuscript with input from all the authors.
\paragraph*{Competing interests}
The authors declare no competing interests.
\paragraph*{Data and material availability}
All data will be available on the Swedish National Data Service.

%

\clearpage

\begin{figure*}[htb!]
\includegraphics[width=\textwidth]{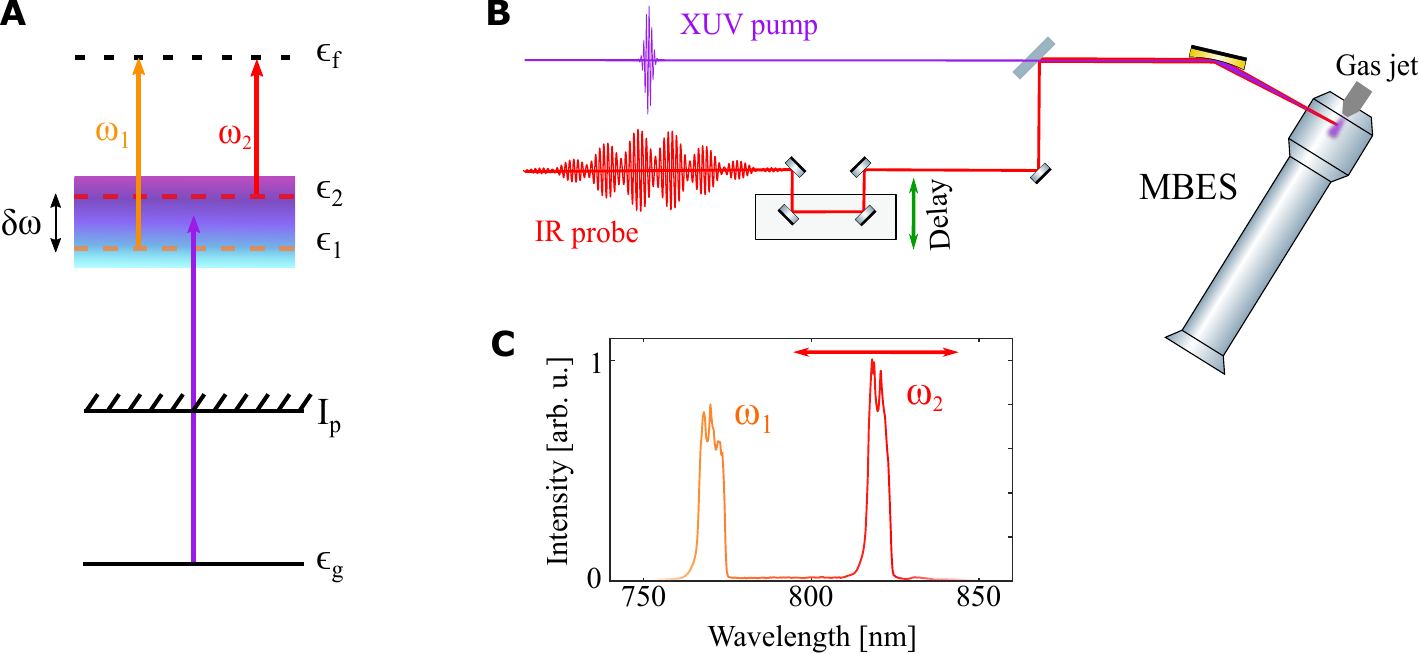}
    \centering
    \caption{Principle of the KRAKEN technique. A) Energy diagram of the KRAKEN scheme. A short XUV pulse ionizes the target (purple arrow) and populates a broad superposition of continuum states. Additional absorption of a bichromatic IR photon with frequency components $\omega_1$ (orange arrow) and $\omega_2$ (red arrow) to a final continuum state with energy $\epsilon_f$ induces interference between intermediate continuum states with energy $\epsilon_1$ and $\epsilon_2$. $\epsilon_g$ and $\textrm{I}_\textrm{p}$ indicate respectively the ground state and the ionization threshold. B) Schematic representation of the experimental setup. An ultrashort XUV pulse and a delayed bichromatic IR pulse are combined and focused in an atomic gas jet. The experiments rely on measuring the kinetic energy of the photoelectrons generated from the interaction with the XUV and IR pulses using a magnetic bottle electron spectrometer (MBES), as a function of the delay between the XUV and the bichromatic probe. C) Example of a bichromatic IR pulse used in the experiment. The red arrow indicates that only the frequency $\omega_2$ is changed in the different measurements, while the frequency $\omega_1$ is fixed.}
    \label{fig1}
\end{figure*}

\begin{figure}[tb!]
    \includegraphics[width=1\textwidth]{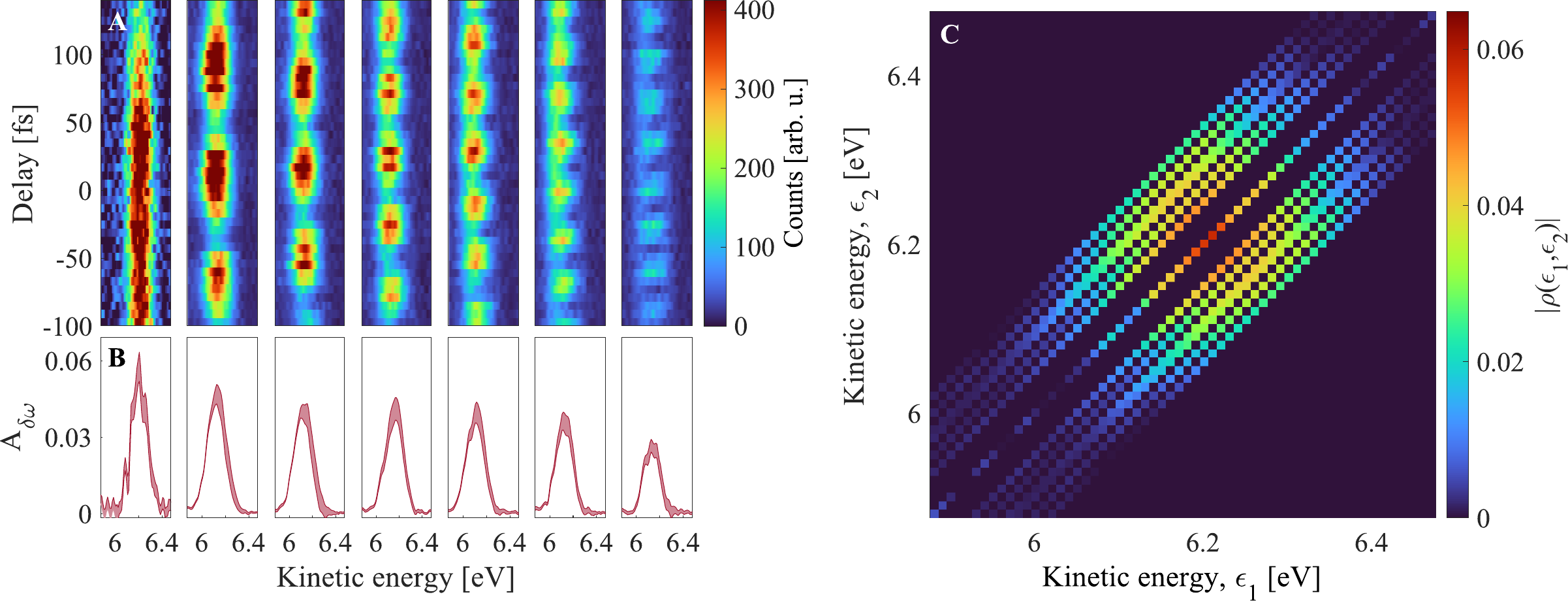}
    \centering
    \caption{Experimental results obtained with the KRAKEN protocol in the case of photoionization of helium atoms. A) Photoelectron spectrograms are acquired for different values of $\delta\omega$ (from left to right: $\hbar\delta\omega=0,41,61,80,98,117,134$ meV). B) Energy-resolved oscillation amplitude $A_{\delta\omega}$ for the different spectrograms. The shaded area indicates the uncertainty of the fit (one standard deviation).  C) Density matrix obtained by inserting the oscillation amplitudes for each $\delta\omega$ at the corresponding position in an initially empty density matrix. Dark blue areas correspond to regions of the density matrix that are not reconstructed.}
    \label{Fig2}
\end{figure}

\begin{figure}[tb!]
    \includegraphics[width=0.8\textwidth]{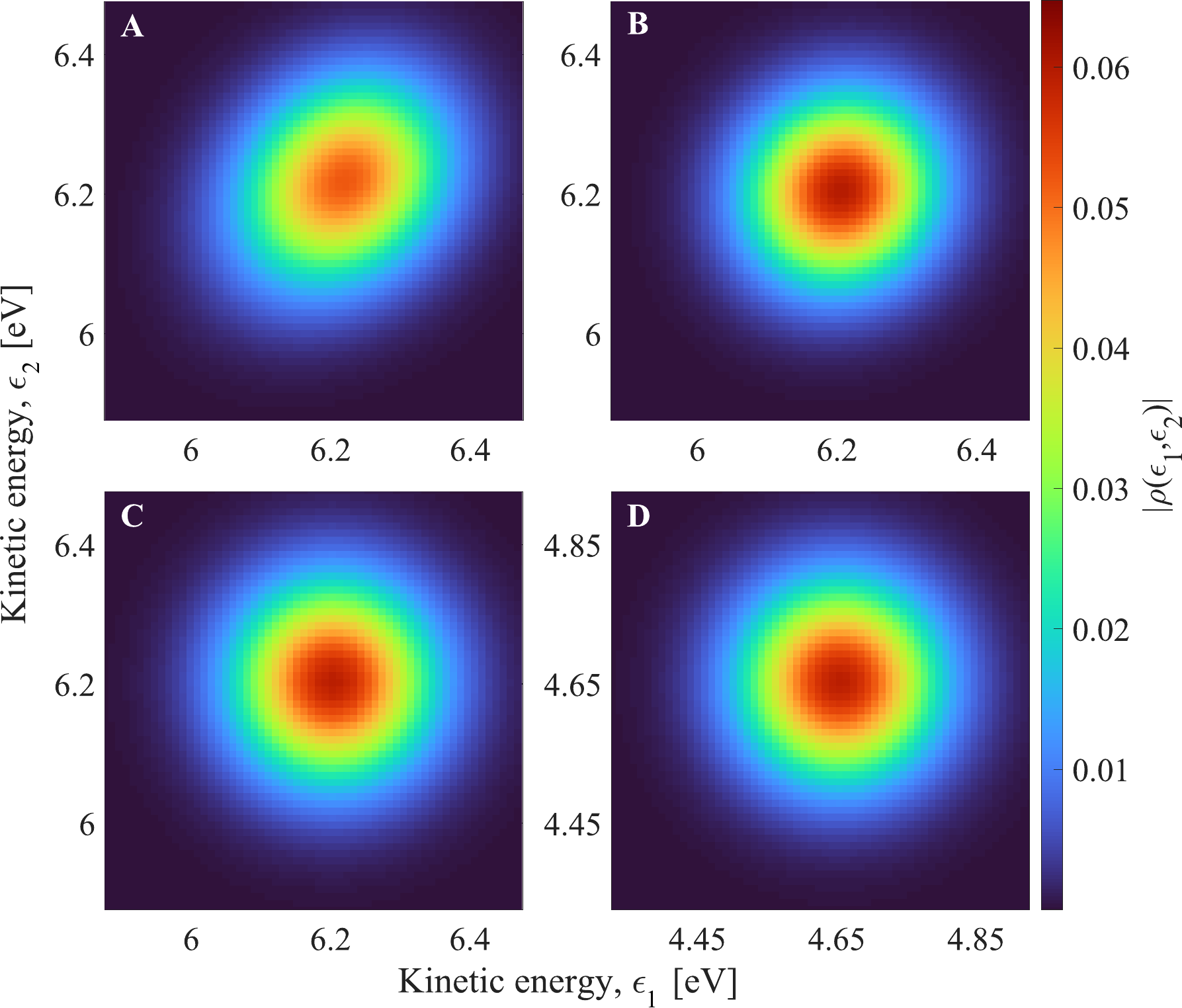}
    \centering
    \caption{Estimated and calculated photoelectron quantum state in helium. A) Density matrix reconstructed from the experimental data. B)  Density matrix reconstructed from the experimental data, accounting for the spectrometer response function in the retrieval algorithm. C) Theoretical density matrix retrieved using the KRAKEN technique. D) Direct theoretical calculation of the photoelectron density matrix in one-photon ionization. The colorbar scale is common to all the figures.}
    \label{Fig3}
\end{figure}

\begin{figure}[tb!]
    \includegraphics[width=0.8\textwidth]{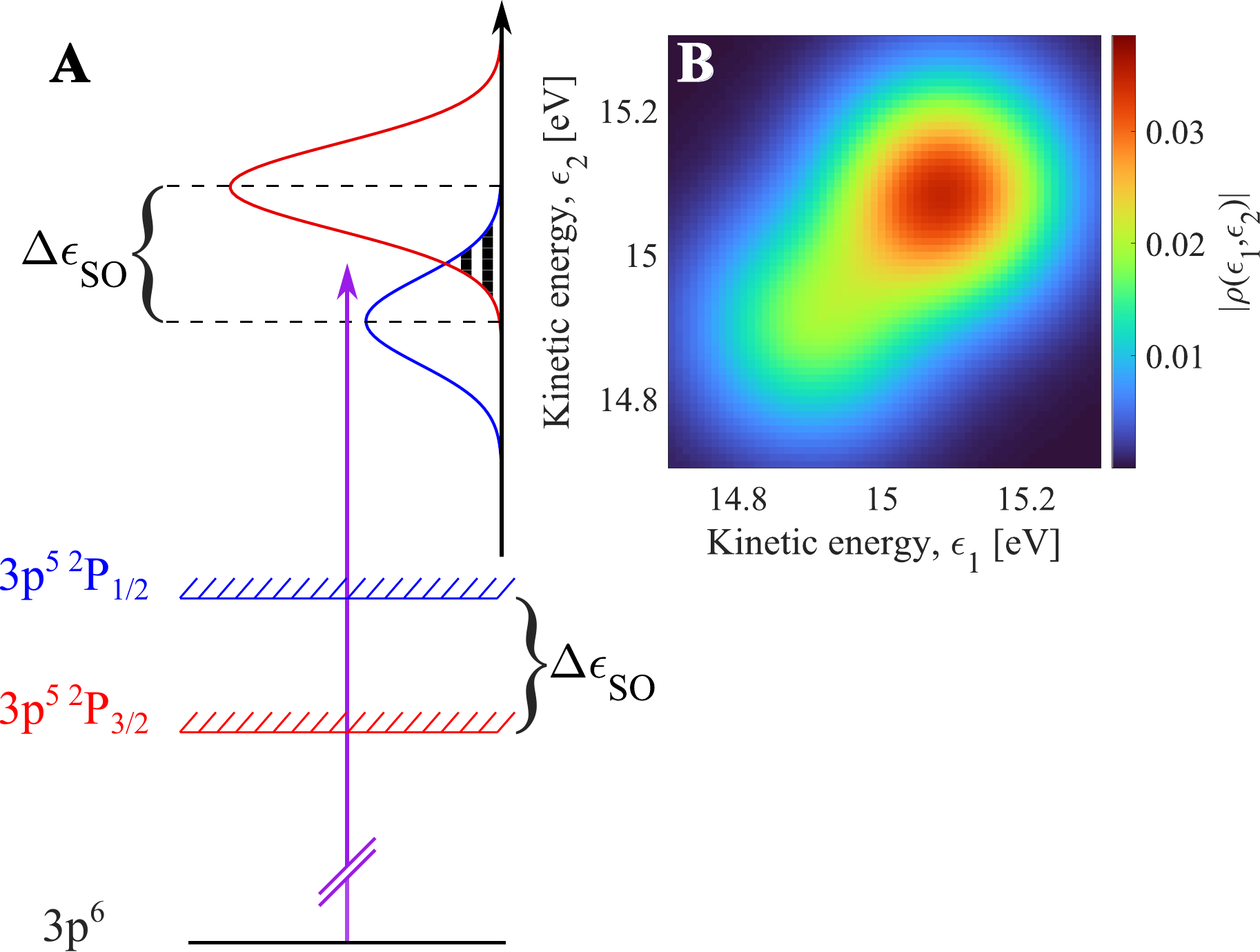}
    \centering
    \caption{Effect of spin-orbit interaction on the photoelectron quantum state in argon. A) Schematic representation of the ionization process. Absorption of an XUV photon promotes an electron from the ground state of a neutral argon atom ($3\mathrm{p}^6$) to the continuum, ionizing the atom. Spin-orbit interaction, splits the ionic ground state  into two mutually orthogonal ionic states, $3\mathrm{p}^5~^2\mathrm{P}_{3/2}$ (red) and $3\mathrm{p}^5~^2\mathrm{P}_{1/2}$ (blue), separated in energy by $\Delta\epsilon_\textrm{so}$. Due to the conservation of energy, the resulting photoelectron spectrum is composed of two peaks, shown in red and blue, associated to the $^2\mathrm{P}_{3/2}$ and $^2\mathrm{P}_{1/2}$ states respectively, and spaced in energy by $\Delta\epsilon_\textrm{so}$. The shaded area indicates the spectral region where the two peaks overlap. B) Experimentally reconstructed quantum state of a photoelectron emitted from an argon atom. }
    \label{Fig4}
\end{figure}

\end{document}